\documentclass[a4paper,12pt]{article}

\usepackage{latexsym,amsmath,amsfonts,amssymb}
\usepackage{multirow}
\usepackage[dvips]{graphicx}
\usepackage{mathrsfs,mathtools}
\usepackage{pstricks,pst-node,pst-text,pst-3d}
\usepackage{slashed}
\usepackage{verbatim}
\usepackage{feynmp}
\usepackage[nosort]{cite}
\usepackage[bulletsep]{collref}

\newcommand{\tr}[1]{\text{Tr}\big[#1\big]}

\newcommand{\sect}[1]{\setcounter{equation}{0}\section{#1}}
\newcommand{\cZ}{{\mathcal Z}}
\newcommand\nn\nonumber
\newcommand\spinchain[2]{\left \{ \begin{matrix} #1 \cr #2 \end{matrix}\right \} }

\renewcommand{\title}[1]{\vbox{\center\LARGE{#1}}\vspace{5mm}}
\renewcommand{\author}[1]{\vbox{\center#1}\vspace{5mm}}
\newcommand{\address}[1]{\vbox{\center\em#1}}
\newcommand{\email}[1]{\vbox{\center\tt#1}\vspace{5mm}}

\begin{document}

\begin{titlepage}
\begin{center}
{\footnotesize \hfill {\tt HU-EP-11/25}}\\
{\footnotesize \hfill {\tt QMUL-PH-11-09 }}\\
{\footnotesize  \hfill {\tt NSF-KITP-11-101}}\\
\vspace{2.5mm}

\title{\sc The two-loop dilatation operator of $\mathcal{N}=4$ super Yang-Mills theory
in the $SO(6)$ sector}
\author{\large George Georgiou${}^{1}$,  Valeria Gili${}^{2}$ and Jan Plef\/ka${}^{3,4}$}
\address{${}^{1}$Demokritos National Research Center, Institute of Nuclear Physics\\
Ag.~Paraskevi, GR-15310 Athens, Greece\\[0.5cm]
${}^{2}$Centre for Research in String Theory,
School of Physics,\\
Queen Mary University of London \\
Mile End Road, London, E14NS, United Kingdom \\[0.5cm]
${}^{3}$Institut f\"ur Physik, Humboldt-Universit\"at zu Berlin,\\
Newtonstra{\ss}e 15, D-12489 Berlin, Germany\\[0.5cm]
${}^{4}$Kavli Institute for Theoretical Physics\\ University of California,
Santa Barbara, CA 93106 - 4030, USA\\[0.5cm]
}

\email{georgiou@inp.demokritos.gr,
v.gili@qmul.ac.uk,\\ plefka@physik.hu-berlin.de}

\end{center}

\abstract{
\noindent
The dilatation operator of planar $\mathcal{N}=4$ super Yang-Mills in the pure scalar
$SO(6)$ sector is derived at the two-loop order. Representation theory allows for eight
free coefficients in an ansatz for the corresponding spin-chain hamiltonian acting
on three adjacent scalar states. While four out
of these follow from the known $SU(2|3)$ sector two-loop dilatation operator, the remaining four
coefficients are derived by diagrammatic techniques and a match to the known
dimension of a length three primary operator. Finally, comments upon the use of this
result for the evaluation of three-point structure functions of scalar operators at the one-loop order are given.
}

\end{titlepage}

\sect{Introduction and Conclusions}\label{sec:intro}

The dilatation operator of planar $\mathcal{N}=4$ super Yang-Mills (SYM)
determines the form of two-point functions of single-trace gauge invariant operators
via its eigenvalues and eigenstates
\cite{Beisert:2003tq}. It lies at the heart of the
integrability of the theory \cite{Beisert:2010jr}. Indeed, the
seminal observation of ref.~\cite{Minahan:2002ve}
 that the one-loop $SO(6)$ dilatation operator is nothing but the
Hamiltonian of an integrable $SO(6)$ spin chain with nearest neighbor interactions (see \cite{Rej:2010ju}
for a recent review) gave rise to
the field of AdS/CFT 
integrability\footnote{Indeed first hints of integrable structures in planar QCD appeared already prior to 
this in the study of high-energy scattering processes \cite{Lipatov:1993yb,Faddeev:1994zg} (see \cite{Belitsky:2004cz,Korchemsky:2010kj} for recent reviews).}.
In an intuitive graphical notation the one-loop $SO(6)$ dilatation operator takes the simple form
\begin{equation}
\label{ham2}
H_{2} =\frac{g_{\text{YM}}^{2}N}{8\pi^{2}}\, \Bigl (\,
\raisebox{-0.3cm}{\includegraphics[width=0.2\textwidth]{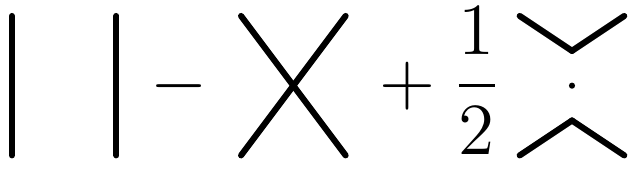}}
\Bigr )
\end{equation}
and acts on nearest neighbor vector states $|\Phi_{a}\Phi_{b}\rangle$ with $\Phi_{a}$ ($a=1,\ldots, 6$)
denoting the six real scalar fields of $\mathcal{N}=4$ SYM. The symbols
correspond to an identity, a permutation and a trace-interaction on $|\Phi_{a}\Phi_{b}\rangle$,
respectively.
This is indeed the only allowed structure
for an integrable nearest-neighbor interaction.
The generalization of \eqref{ham2} to the full
$SU(2,2|4)$ excitation spectrum of the theory at the one-loop order
turns out to preserve the integrable structure \cite{Beisert:2003yb}. In the
higher loop corrections to the dilatation operator
 the spread of the spin-interactions grows linearly with the loop-order.
Then the initially pure scalar operators in the
$SO(6)$ sector start to mix with operators including bi-fermion and covariant derivative insertions,
which is why an explicit construction of the two-loop generalization to \eqref{ham2} had not been performed
to date \footnote{The $SO(6)$ sector in the thermodynamic limit was studied in \cite{Minahan:2004ds,Beisert:2004ag}.}.
The situation is different for the closed sectors of $SU(2)$ spanned by two complex
scalar fields $Z=\Phi_{1}+i\Phi_{2}$ and $Z_{1}=\Phi_{3}+i\Phi_{4}$ itself being embedded in
 the maximally closed compact subsector
of $SU(2|3)$ involving in addition the third complex scalar $Z_{2}=\Phi_{5}+i\Phi_{6}$ and
gluino field $\Psi^{4}_{\alpha}$\, ($\alpha=1,2$).
In this closed sector the form of the dilatation operator is entirely determined to the three loop order by the
$SU(2|3)$ symmetry, the topology of underlying Feynman graphs, the protectedness of $\frac{1}{2}$ BPS
states and the existence of BMN scaling\footnote{As a matter of fact $\mathcal{N}=4$ SYM is known to
violate BMN scaling starting at the four loop level.} \cite{Beisert:2003ys}. 
This result was confirmed by an explicit three-loop calculation in \cite{Sieg:2010tz}
Note that it is the trace-term
in \eqref{ham2} which does not couple to these closed subsectors as the contraction of any two scalars from the
set $\{Z,Z_{1},Z_{2}\}$ vanishes. Similarly in the important non-compact $SL(2)$ subsector
comprised of $Z$ and covariant derivatives acting on it, the dilatation operator has been
perturbatively constructed to
the two-loop order \cite{Eden:2005bt}, see also \cite{Eden:2005ta} for partial results
at three-loops. For an overview on the status of perturbative constructions
of the $\mathcal{N}=4$ SYM dilatation operator see \cite{Sieg:2010jt}.

The recent progress in the understanding of the spectral problem of $\mathcal{N}=4$ SYM
was based on the powerful \emph{assumption} of integrability of the underlying all-loop dilatation operator,
i.e.~its two-particle scattering factorization property, and constructing the underlying S-matrix
essentially through its $SU(2,2|4)$ symmetry structure (see \cite{Beisert:2010jr} for reviews).
However, the exact perturbative expression of the dilatation operator remains
unknown to date. This implies
that while the Bethe-ansatz techniques (and its generalizations) give us the
eigenvalues of the dilatation operator, the form of the eigenstates for the resolution of 
the hard problem of operator mixing in superconformal theories \cite{Belitsky:2007jp}
remains unknown. This is particularly unfortunate, as these eigenstates enter crucially
in the determination of the three-point functions in any perturbative study of the latter.

This necessity was the motivation for this work to establish the form of the two-loop dilatation operator in
the pure scalar $SO(6)$ sector generalizing \eqref{ham2}. Our central result takes
the diagramatic form
\begin{equation}
\label{ham4}
\raisebox{-0.3cm}{\includegraphics[width=0.9\textwidth]{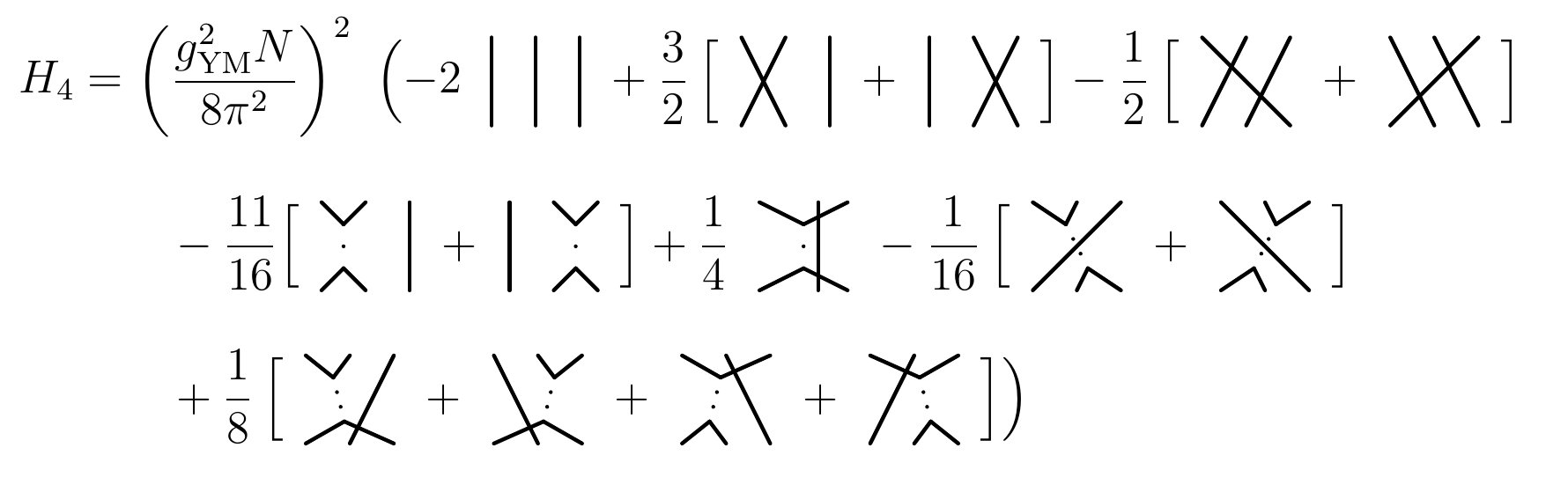}}
\end{equation}
where the first four terms already follow from the $SU(2|3)$ result of \cite{Beisert:2003ys}.
We derived this by classifying the possible terms in a  general ansatz and by performing a minimal number of explicit
Feynman diagrammatic computations along with requiring a match with a known two-loop scaling dimension of a
length three-state.

As mentioned above this result will be of crucial use for the computation of three-point functions
at the one-loop order. Three-point functions of single-trace operators in $\mathcal{N}=4$ SYM were
studied at weak \cite{Beisert:2002bb,Roiban:2004va,Okuyama:2004bd,Alday:2005nd,Alday:2005kq,Georgiou:2009tp} and recently also at strong-coupling
\cite{Zarembo:2010rr,Costa:2010rz,Roiban:2010fe,Hernandez:2010tg,Georgiou:2010an,Park:2010vs,Bak:2011yy,Bissi:2011dc,Hernandez:2011up,Ahn:2011zg}
as well as from the integrability perspective \cite{Escobedo:2010xs,Escobedo:2011xw}
(see also \cite{Kristjansen:2010kg} for a brief review).
Our result will be instrumental for  curing the missing mixing contributions of
the preprint \cite{Grossardt:2010xq} which will appear in a companion paper \cite{us2}.
In addition, the above Hamiltonian represents the pure scalar piece of the general $SU(2,2|4)$
dilatation operator at the two-loop level. It would be interesting to see to what extent the form
of \eqref{ham4} together with the rich symmetry (and integrability) structure of the
full theory already determine the complete two-loop dilatation operator of $\mathcal{N}=4$ SYM.

\section{The 2-loop planar $SO(6)$ dilatation operator}
\label{sec:Hamiltonian}
In this section, we 
evaluate the pure scalar $SO(6)$  piece of the two-loop 
dilatation operator $H_4$ of $\mathcal{N}=4$ SYM, i.e.~the part of the 
dilatation operator which maps three scalars
to three scalars.
Although the  $SO(6)$ sector of $\mathcal{N}=4$ SYM is not closed beyond one-loop
due to the mixing with of three scalars with two fermions or one scalar with
a covariant derivative insertion,
the considered part of the dilatation operator is of interest:
On the one hand it is a further step towards the construction of the explicit form of the
 complete $SU(2,2|4)$ two-loop dilatation operator of the theory.
On the other hand the knowledge of this operator is necessary
in order to resolve the mixing among primary operators in long multiplets up to order $g^2$
relevant for the determination of structure functions in three-point correlators at 
one-loop.

After these comments we proceed to the calculation. The two-loop dilatation
operator or $SO(6)$ spin-chain Hamiltonian $H_4$ can act
on three letters the most. Each of these letters transform in the ${\bold 6}$
irreducible representation (irrep) of $SU(4)\sim SO(6)$. The tensor product of three ${\bold 6}$ -dimensional representations of $SU(4)$
can be decomposed in the following eight irreps
\begin{eqnarray}\label{decomposition}
(1,1,0)_{\bold 6}\otimes(1,1,0)_{\bold 6}\otimes(1,1,0)_{\bold 6}=(1,1,0)_{\bold 6} \oplus (2,2,0)_{\bold 5 \bold 0}\oplus (3,2,1)_{\bold 6 \bold 4}\nn \\
\oplus(1,1,0)_{\bold 6}\oplus (3,2,1)_{\bold 6 \bold 4}\oplus (2,0,0)_{\bold 1 \bold 0}\oplus (2,2,2)_{\bold 1\bold 0}\oplus(1,1,0)_{\bold 6}.
\end{eqnarray}
Here $(f_1,f_2,f_3)_{\bold d}$ denotes a Young tableau of $SU(4)$ with $f_1$ boxes in the first row, $f_2$ boxes in the second row
and $f_3$ boxes in the third row, while ${\bold d}$ is the dimension of the representation.
By taking into account the fact that the Hamiltonian should be a singlet under $SU(4)$ one can
write the most general form for $H_4$. This reads
\begin{eqnarray}\label{H4general}
H_4= \sum_{i=1}^{8} d_i P_i
\end{eqnarray}
where $P_i$ is the projector of each of the irreps appearing in \eqref{decomposition}.
From \eqref{H4general} it is apparent that to fully determine $H_4$ one has to find
the eight coefficients $d_i$.

\begin{figure}[!t]
\centering
 \includegraphics[width=0.999\textwidth]{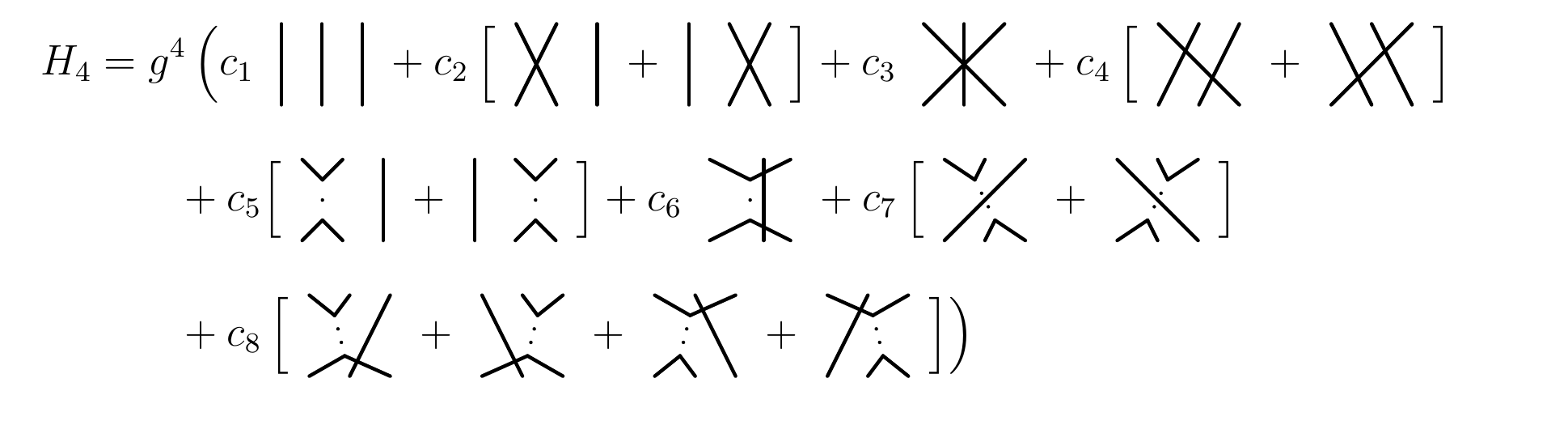}
  \caption{Graphical representation of the $SO(6)$ dilatation operator $H_{4}$ of \eqref{ansatz}. The
  first line is the $SU(2|3)$ bosonic piece of Beisert \cite{Beisert:2003ys}, the central result of this
  work is the determination of the remaining coefficients $c_{5},c_{6},c_{7}$ and $c_{8}$.}
  \label{Fig:Ham}
\end{figure}

In what follows, it will be more convenient to parametrize the Hamiltonian
in a different but equivalent way. Namely we write it as,
\begin{align}\label{ansatz}
H_{4}&= g^4\, \Bigl (
c_1\, \spinchain{abc}{abc}
+ c_2\, \Bigl [ \spinchain{bac}{abc} +  \spinchain{acb}{abc}\Bigr ]
+ c_3 \, \spinchain{abc}{cba}
+c_4 \, \Bigl [ \spinchain{bca}{abc} +  \spinchain{cab}{abc}\Bigr ] \nn\\
& \qquad +c_5 \Bigl [ \spinchain{bbc}{aac} +  \spinchain{cbb}{caa}\Bigr ]
+ c_6 \, \spinchain{bcb}{aca}
+c_7 \, \Bigl [ \spinchain{bbc}{caa} +  \spinchain{cbb}{aac}\Bigr ]\nn\\
& \qquad +c_8\, \Bigl [ \spinchain{bbc}{aca} +  \spinchain{cbb}{aca} \Bigr ]
+c_8^{\ast}\, \Bigl [\spinchain{bcb}{aac} +  \spinchain{bcb}{caa} \Bigr ]\Bigr )
\, , \qquad g^2:=\frac{g_{YM}^2 N}{8 \pi^2}\, ,
\end{align}
see figure \ref{Fig:Ham} for a graphical representation.

The action of these operators on the three scalar letters is defined as follows
\begin{equation}\label{def}
\spinchain{abc}{def}\, |\Phi_{i}\,\Phi_{j}\Phi_{k}\rangle =
\delta_{di}\, \delta_{ej}\, \delta_{fk}\, |\Phi_{a}\,\Phi_{b}\Phi_{c}\rangle\, ,
\end{equation}
where repeated indices are summed over. In \eqref{def} $a,b,c,\ldots=1,\ldots,6$ denote the $SO(6)$ indices
of the scalar fields $\Phi_{i}$.
In writing \eqref{ansatz} we have made use of the invariance of $H_4$ under parity
(see eq. 2.154 and 2.155 of \cite{Beisert:2004ry})
\begin{eqnarray}\label{parity}
{\cal P}H_4{\cal P}^{-1}=H_4, \, \, \, \,{\cal P}| A_1 A_2 A_3\rangle=-| A_3 A_2 A_1\rangle
\end{eqnarray}
to make the coefficients of the operators in each of the brackets of \eqref{ansatz} equal.
Furthermore, the hermiticity of $H_4$ imposes the condition that the two coefficients
appearing in the last line of \eqref{ansatz} should be the complex conjugate of each other.
However, the analysis of the corresponding diagrams that will be performed
shows that these coefficients are real and, as a consequence, equal.
We conclude that the full scalar 2-loop Hamiltonian $H_4$ can be written in terms of 8 independent
constants $c_i,\, i=1,...,8$.
In what follows, 4 of those constants will be determined by applying \eqref{ansatz}
to states in a $SU(2|3)$ subsector for which the Hamiltonian is known
from the work of \cite{Beisert:2003ys} , 3 from Feynman diagrams and the last one
by comparison to the known 2-loop anomalous dimension of the length 3 operator $\tr{\Phi_i \Phi_i \Phi_j}$.

We start by considering the action of  $H_4$ on the $SU(2|3)$ state
$| Z_1 Z_2 Z\rangle$\footnote{Recall the complex combinations
$Z=\Phi_{1}+i\Phi_{2}$, $Z_{1}=\Phi_{3}+i\Phi_{4}$ and
$Z_{2}=\Phi_{5}+i\Phi_{6}$.}. These three complex scalar fields together with $\psi^4_{\alpha}$
form the basis of a closed $SU(2|3)$ subsector of the full $PSU(2,2|4)$ algebra.
The supercharges that close the sub-algebra are ${\bar Q}^1_{\dot \alpha}\, ,{\bar Q}^2_{\dot \alpha}\, ,{\bar Q}^3_{\dot \alpha}$
and ${\bar S}_{1 \dot \alpha}\, ,{\bar S}_{2 \dot \alpha}\, ,{\bar S}_{3 \dot \alpha}$.
Here and in the rest of this note we use the conventions of \cite{Georgiou:2008vk}.
When written in terms of the six real scalars $\Phi_i, \, i=1,...,6$ the state $| Z_1 Z_2 Z\rangle$ becomes
a sum of terms none of which has repeated $SO(6)$ indices. This means that only the first
four terms in the Hamiltonian $H_4$ will give non-zero contributions. Thus, one gets
\begin{eqnarray}\label{ZZ1Z2}
H_4| Z_1 Z_2 Z\rangle=c_1 g^4| Z_1 Z_2 Z\rangle+c_2 g^4 (| Z_2 Z_1 Z \rangle+| Z_1 Z Z_2 \rangle)+\nn \\
c_3 g^4| Z Z_2 Z_1 \rangle
+c_4 g^4(| Z_2 Z Z_1\rangle+| Z Z_1 Z_2\rangle).
\end{eqnarray}
This should be compared to the result obtained in \cite{Beisert:2003ys} where the 2-loop
Hamiltonian of the $SU(2|3)$ subsector was derived by exploiting the algebra.
\begin{eqnarray}\label{SUBeisert}
H_4| Z_1 Z_2 Z\rangle=-2 g^4 | Z_1 Z_2 Z\rangle+\frac{3}{2} g^4(| Z_2 Z_1 Z \rangle+| Z_1 Z Z_2 \rangle)
-\frac{1}{2} g^4(| Z_2 Z Z_1\rangle+| Z Z_1 Z_2\rangle).
\end{eqnarray}
Direct comparison of \eqref{ZZ1Z2} and \eqref{SUBeisert} gives
\begin{eqnarray}\label{c1to4}
c_1=-2, \qquad c_2=\frac{3}{2}, \qquad c_3=0, \qquad c_4=-\frac{1}{2}.
\end{eqnarray}

Before evaluating the rest of the unknown coefficients, let us briefly review the construction of the anomalous dimension matrix.
This matrix can be calculated from the mixing matrix
$\cZ^A_{\, \, \,B}$ via
\begin{eqnarray}\label{anom.matrix}
\gamma=\mu \frac{\partial {\log \cZ}}{\partial \mu}|_{\lambda_b}= \mu \frac{\partial \cZ}{\partial \mu}\cZ^{-1}|_{\lambda_b}\, ,
\end{eqnarray}
where $\mu$ is the renormalisation scale of the theory while the mixing matrix
is determined by demanding the
finiteness of the correlation function
\begin{eqnarray}\label{finitecorr.}
\langle {\cal O}^{\dagger}_{B \, ren}(x_2) \, \, \, {\cal O}^A_{ren}(x_1)\rangle .
\end{eqnarray}
Here  ${\cal O}^A_{ren}(x)$ denotes the renormalized
operator which is given in terms of the bare operators ${\cal O}^B$ by the relation
\begin{eqnarray}\label{renorm1}
 {\cal O}^A_{ren}=\cZ^A_{\, \, \,B}{\cal O}^B .
\end{eqnarray}
One can employ renormalized perturbation theory by adding the appropriate
counter-terms to the $\mathcal{N}=4$ SYM action using
supersymmetric regularization by
dimensional reduction. Expressing the composite operators in \eqref{renorm1} in terms of the
renormalized fields $\Phi_{i \, ren}=Z_{\Phi}^{1/2}\Phi_{i}$ we get
\begin{eqnarray}\label{renorm2}
 {\cal O}^A_{ren}=\cZ^A_{\, \, \,B}{\cal O}^B=\cZ^A_{\, \, \,B}Z_{\Phi}^{-L/2}{\tilde{\cal O}}^B={\tilde \cZ}^A_{\, \, \,B} {\tilde{\cal O}}^B,
 \, \, \,\, \, \,{\tilde{\cal O}}^B=\tr{\Phi_{i_1 \, ren}...\Phi_{i_L \, ren}}
\end{eqnarray}

The strategy to determine the anomalous dimension matrix $\gamma$ should be clear.
One substitutes \eqref{renorm2} in \eqref{finitecorr.} and determines ${\tilde \cZ}^A_{\, \, \,B}$
by demanding that the correlator involving the renormalized operators is finite.
Then one uses the relation
\begin{eqnarray}\label{Zexp}
\cZ^A_{\, \, \,B}={\tilde \cZ}^A_{\, \, \,B}Z_{\Phi}^{L/2}=1+\cZ^{A}_{1\, \, \,B}+\cZ^{A}_{2\, \, \,B}+...
\end{eqnarray}
to determine $\cZ^A_{\, \, \,B}$. In \eqref{Zexp} $\cZ^{A}_{1\, \, \,B}$ and 
$\cZ^{A}_{2\, \, \,B}$ denote
the $g^2$ and $g^4$ contributions to $\cZ$, respectively.
One can then plug \eqref{Zexp} in \eqref{anom.matrix} to get for the 
two-loop anomalous dimension matrix
\begin{eqnarray}\label{2-loopanom.matrix}
\gamma^{A}_{2\, \, \,B}=\mu \frac{\partial ({\cZ^{A}_{2\, \, \,B}-\frac{1}{2}\cZ^{A}_{1\, \, \,C}\cZ^{C}_{1\, \, \,B}})}{\partial \mu}|_{\lambda_b}.
\end{eqnarray}
Since it is a 2-loop contribution $\cZ^{A}_{2\, \, \,B}$ will have poles of order $1/\epsilon^2$ and $1/\epsilon$.
However, the $1/\epsilon^2$ pole should cancel against the $1/\epsilon^2$ pole coming from the
$-\frac{1}{2}\cZ^{A}_{1\, \, \,C}\cZ^{C}_{1\, \, \,B}$ term, resulting in a well-defined anomalous dimension matrix
in the limit $\epsilon \rightarrow 0$.
It is instructive to expand the 2-point correlator of the renormalized operators as follows:
\begin{eqnarray}\label{finitecorr.2}
&&\langle {\cal O}^{\dagger}_{B \, ren}(x_2) \, \, \, {\cal O}^A_{ren}(x_1)\rangle=
\langle {\tilde{\cal O}}^{\dagger}_{D}(x_2){\tilde \cZ}^D_{\, \, \,B} \, \, \, {\tilde \cZ}^A_{\, \, \,C} {\tilde{\cal O}}^C (x_1)\rangle= \nonumber\\
&&{\tilde \cZ}^B_{\, \, \,B}\Big( {\tilde \cZ}^A_{\, \, \,A} \langle \tilde {\cal O}^{\dagger}_{B}(x_2) \, \, \, \tilde {\cal O}^A(x_1)\rangle
+{\tilde \cZ}^A_{\, \, \,B} \langle \tilde {\cal O}^{\dagger}_{B}(x_2) \, \, \, \tilde {\cal O}^B(x_1)\rangle+
\sum_{C\neq A,B}{\tilde \cZ}^A_{\, \, \,C} \langle \tilde {\cal O}^{\dagger}_{B}(x_2) \, \, \, \tilde {\cal O}^C(x_1)\rangle\Big)+ \nonumber\\
&&{\tilde \cZ}^A_{\, \, \,A}\Big( {\tilde \cZ}^A_{\, \, \,B} \langle \tilde {\cal O}^{\dagger}_{A}(x_2) \, \, \, \tilde {\cal O}^A(x_1)\rangle
+\sum_{C\neq A,B}{\tilde \cZ}^C_{\, \, \,B} \langle \tilde {\cal O}^{\dagger}_{C}(x_2) \, \, \, \tilde {\cal O}^A(x_1)\rangle\Big)
+\nonumber\\ && \sum_{C,D\neq A,B}{\tilde \cZ}^D_{\, \, \,B}{\tilde \cZ}^A_{\, \, \,C}
\langle \tilde {\cal O}^{\dagger}_{D}(x_2) \, \, \, \tilde {\cal O}^C(x_1)\rangle.
\end{eqnarray}
In the equality of the first line we have used the fact that in the scalar sector ${\tilde \cZ}$ can be chosen
to be real and symmetric, i.e. ${\tilde \cZ}^{\dagger}={\tilde \cZ}$.
\begin{figure}[!t]
  \centering
  \includegraphics[width=0.7\textwidth]{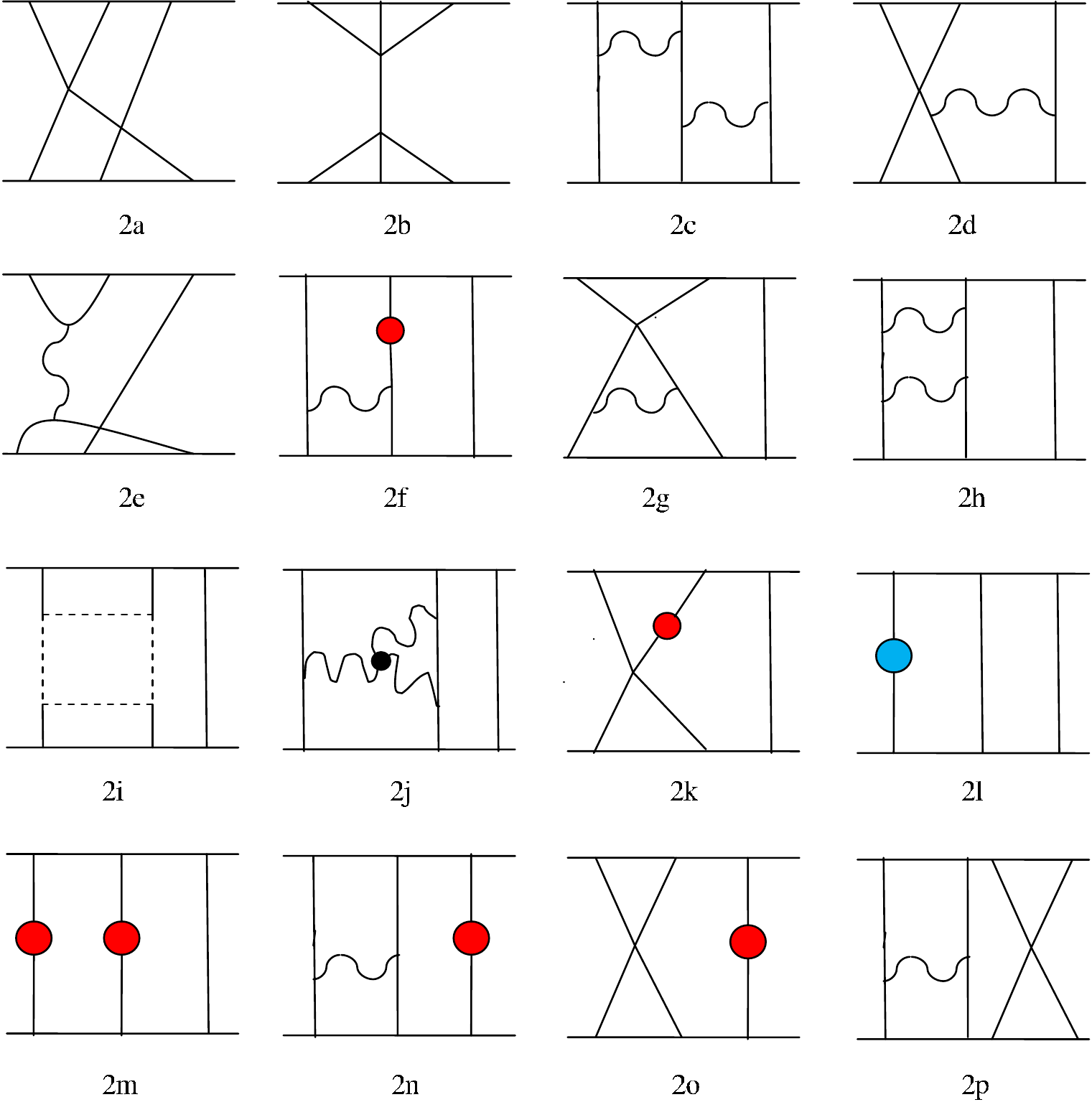}
  \caption{Diagrams contributing to the correlator \eqref{finitecorr.2} at 2-loops.
  The solid, wiggly and  dashed
lines represent scalars, gluons and fermions, respectively. The red blob denotes the 1-loop correction to the scalar
propagator while the blue blob of 1l the 2-loop correction.
  In 2a-2e
 the interactions connect three pairs of letters. In 2f-2k they connect two pairs of letters
 while in 2l just one pair. Finally,  2m-2p are the disconnected
 diagrams which will give the $\log^2$ term necessary for the conformal structure of the correlator.
 We should mention that we have not drawn any diagrams coming from the counter-terms added to the Lagrangian and that
 diagram 2e is 0. }
  \label{fig:alldiagrams}
\end{figure}

In Figure 2 we depict all different kinds of diagrams that can in principle contribute to
the correlator \eqref{finitecorr.2} at 2-loop order. Let us notice that the diagram 1e is 0 since it is proportional
to $(\partial_1-\partial_2)\cdot(\partial_3-\partial_4)H_{12,34}$ with $x_1=x_2$. This last condition is due to the fact that
the propagator of the top scalar that emits the gluon has  both its ends at the same point.
The expression for $H_{12,34}$ can be found in \cite{Beisert:2002bb}.

To illustrate the procedure let us focus on the following anomalous dimension matrix element of the $SU(2|3)$ sector
$H_4|Z_1 Z_2 Z\rangle= -\frac{1}{2}g^4| Z Z_1 Z_2 \rangle+...$. In other words, we have chosen ${\cal O}^A=|Z_1 Z_2 Z\rangle$
and ${\cal O}^B=|Z Z_1 Z_2 \rangle$ which means ${\cal O}^{\dagger}_{B}= \langle {\bar Z}_2{\bar Z}_1{\bar Z}|$.
By inspecting Figure 2 one can convince oneself that there is a single diagram contributing in
${\tilde \cZ}^A_{\, \, \,B} $, shown in Figure 3c. This diagram corresponds to
the first  correlator appearing in the second  line of \eqref{finitecorr.2} and is already of order $g^4$.
A first observation is that the one loop contribution to ${\tilde Z}^A_{\, \, \,B} $ is zero.
This implies that one can set all diagonal $ {\tilde \cZ}^A_{\, \, \,A}$ in \eqref{finitecorr.2} equal to 1.
The second ingredient one needs is the product of two one loop $\cZ$'s appearing both in \eqref{finitecorr.2} and
\eqref{2-loopanom.matrix}, namely $\cZ^{A}_{1\, \, \,C}\cZ^{C}_{1\, \, \,B}$.  The operator ${\cal O}^C$ appearing
in the second, third and fourth line of \eqref{finitecorr.2} can be found by cutting the 2-loop diagram of Figure 3c
leaving one vertex on one side of the cut and the other vertex on the other side.
In this way we obtain ${\cal O}^C=|  Z_1 Z Z_2 \rangle$.
Demanding that \eqref{finitecorr.2} is finite up to order $g^4$ will give $\cZ^A_{2\, \, \,B}$ which can then be plugged in
 \eqref{2-loopanom.matrix} to give the correct 2-loop matrix element, $-\frac{1}{2}g^4$.
This value was obtained in \cite{Beisert:2003ys} by exploiting the algebra of the $SU(2|3)$ subsector.

We now turn to the evaluation of the remaining 4 unknown coefficients.
To start with, we consider the action of the 2-loop Hamiltonian of \eqref{ansatz} on the state $| Z_1 {\bar Z}_1 Z\rangle$.
This gives
\begin{eqnarray}\label{Z1barZ1Z}
H_4| Z_1 {\bar Z}_1 Z\rangle=-2 g^4 | Z_1 {\bar Z}_1 Z\rangle+\frac{3}{2} g^4( | Z_1 Z {\bar Z}_1 \rangle+ |  {\bar Z}_1 Z_1 Z\rangle)
-\frac{1}{2} g^4( | Z Z_1  {\bar Z}_1 \rangle+ |  {\bar Z}_1 Z Z_1\rangle)\nn \\
+c_5g^4(|Z_1 {\bar Z}_1 Z\rangle+|{\bar Z}_1 Z_1 Z\rangle+|Z_2 {\bar Z}_2 Z\rangle+|{\bar Z}_2 Z_2 Z\rangle+|Z {\bar Z} Z\rangle+|{\bar Z} Z Z\rangle)
\nn \\
+c_7 g^4(| Z Z_1 {\bar Z}_1 \rangle+|Z {\bar Z}_1 Z_1 \rangle+| Z Z_2 {\bar Z}_2 \rangle+| Z {\bar Z}_2 Z_2 \rangle+|Z Z {\bar Z} \rangle+| Z {\bar Z} Z\rangle)\nn \\
+c_8 g^4(|Z_1 Z {\bar Z}_1 \rangle+|{\bar Z}_1 Z Z_1 \rangle+|Z_2 Z {\bar Z}_2 \rangle+|{\bar Z}_2 Z Z_2 \rangle+|Z Z {\bar Z}\rangle+|{\bar Z} Z Z\rangle)\nn \\
\end{eqnarray}

Using
$H_4| Z_1 {\bar Z}_1 Z\rangle=
c_7 | Z {\bar Z}_2 Z_2 \rangle+\ldots\,$ ,
where the dots denote terms that are different from the bra $| Z {\bar Z}_2 Z_2 \rangle$,
the matrix element $c_7$ can be determined.
\begin{figure}[!t]
  \centering
  \includegraphics[width=0.6\textwidth]{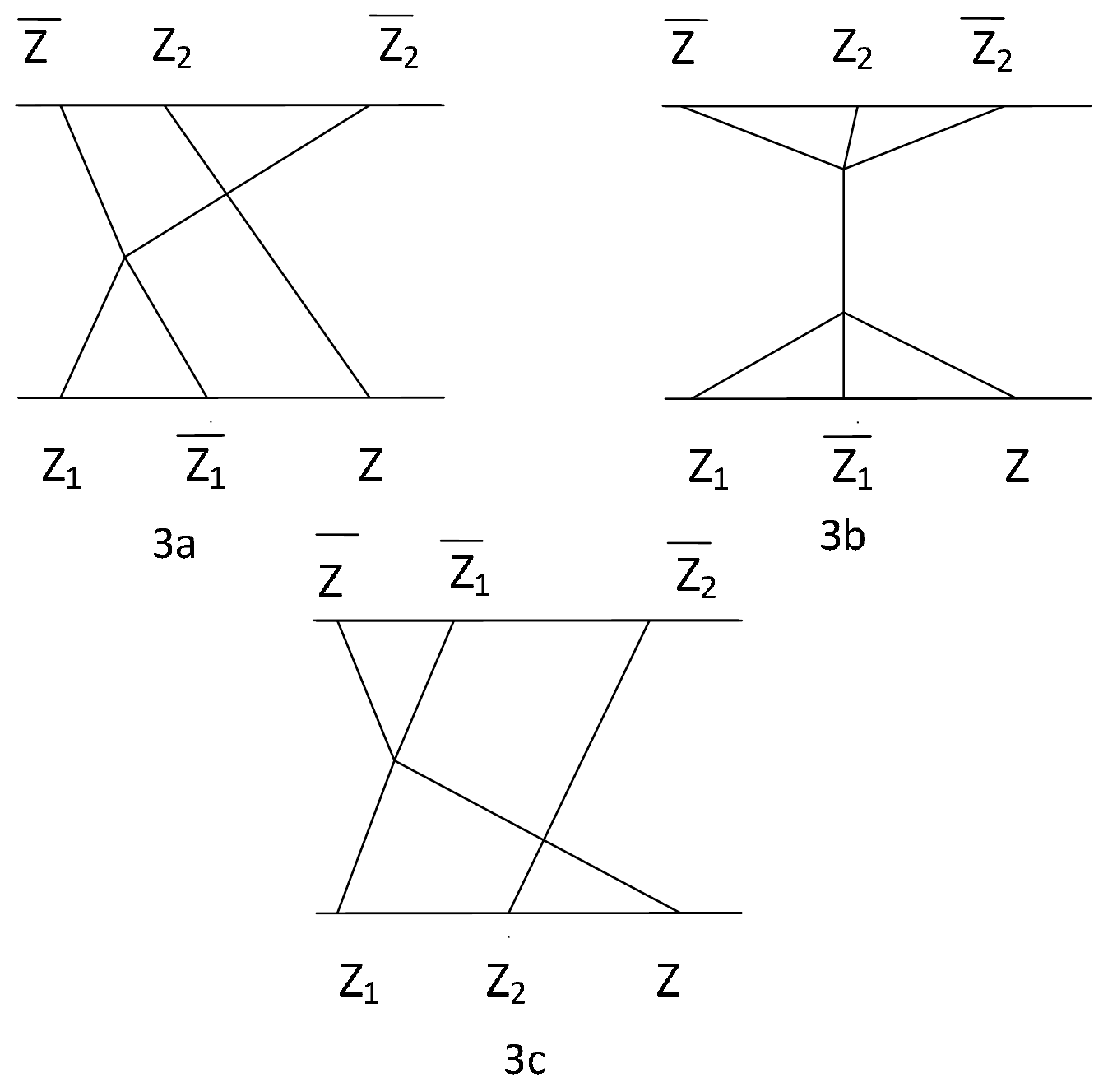}
  \caption{Diagrams 3a and 3b contribute to the coefficient $c_7$. 3c is the analogous diagram for the $SU(2|3)$ sector.
  By inspecting \eqref{4vertex} one can verify that diagram 3a has the same space-time structure as 3c but its value is $1/4$
  of the value of the latter.  }
  \label{fig:current1}
\end{figure}
Because of the fact that the field $Z$ jumps over both $ Z_1$ and ${\bar Z}_1$ the number of diagrams contributing is limited.
In fact there are only two. They are the diagrams 3a and 3b depicted in Figure 3.
We need the contribution of these diagrams to the renormalisation matrix $Z^A_{\, \, \,B}$ in terms of which
the anomalous dimensions matrix $\gamma$ is  calculated.
The contribution of the diagram 3b is calculated in appendix A. The contribution of the diagram 3a
can be deduced from the results of \cite{Beisert:2003ys} as follows.
Consider for a moment the diagram 3c of Figure 3. As mentioned above, this diagram gives the only contribution to the $g^4$
anomalous dimension matrix element $H_4| Z_1 Z_2 Z\rangle= -1/2 g^4 | Z Z_1 Z_2\rangle$.
The crucial observation is that 3c has the same space-time structure as 3a. The only difference comes from the
flavor structure of the two diagrams. One can rewrite the 4-scalar vertex of $\mathcal{N}=4$ SYM as
\begin{eqnarray}\label{4vertex}
 V=g_{YM}^2 \tr{2[Z_i,Z_j][{\bar Z}_i,{\bar Z}_j]-[Z_i,{\bar Z}_i][Z_j,{\bar Z}_j]}=
 2 g_{YM}^2 \tr{2 Z Z_1 {\bar Z} {\bar Z}_1+2Z_1Z{\bar Z}_1{\bar Z}\nonumber\\
 -Z Z_1{\bar Z}_1{\bar Z}-Z_1Z{\bar Z}{\bar Z}_1
 -Z{\bar Z}Z_1{\bar Z}_1-{\bar Z}Z{\bar Z}_1Z_1+...-{\bar Z}Z{\bar Z}Z+{\bar Z}{\bar Z}Z Z+...},
\end{eqnarray}
where the first set of dots denote two terms similar to the one written but with the fields being
$Z_1$, $Z_2$ or $Z$, $Z_2$ instead of $Z$, $Z_1$ while the second set of dots terms where instead of $Z$ one has
$Z_1$ or $Z_2$.
By inspecting \eqref{4vertex}
one can see that the diagram 3c is 4 times the diagram 3a. This is so because each vertex
of 3c is -2 times each vertex of 3a. This means that the divergence of diagram 3a is $1/4$
that of 3c. The same holds for the 1-loop squared term $\cZ^{A}_{1\, \, \,C}\,
\cZ^{C}_{1\, \, \,B}$
since each 1-loop $\cZ^{A}_{1\, \, \,C}$ appearing in $SU(2|3)$ sector process is again -2 times
the corresponding one-loop $\cZ$ contributing in the process we are considering.
Consequently, the logarithmic divergence and thus
the contribution to the anomalous dimension matrix $\gamma$ of the diagram 3a is $1/4$
that of the diagram 3c.
We write this result as
\begin{eqnarray}\label{1a}
H_4^{(3a)}| Z_1 {\bar Z}_1 Z\rangle=\frac{1}{4}\,\,\, \frac{-1}{2}g^4| Z {\bar Z}_2 Z_2\rangle
=-\frac{1}{8}g^4| Z {\bar Z}_2 Z_2\rangle.
\end{eqnarray}
The contribution of diagram 3b to $Z^A_{\, \, \,B}$ is evaluated in Appendix A. It has a simple pole and is given by
\begin{eqnarray}\label{1bZ}
\cZ^{A(3b)}_{\, \, \,B}=\frac{(g_{YM}^2 \mu^{2 \epsilon} N)^2}{32 (8 \pi^2)^2} \frac{1}{2 \epsilon},
\end{eqnarray}
from which one can deduce the contribution to the corresponding  anomalous dimension matrix element to be
\begin{eqnarray}\label{1b}
H_4^{(3b)}| Z_1 {\bar Z}_1 Z\rangle=\frac{1}{16}g^4| Z {\bar Z}_2 Z_2\rangle,
\end{eqnarray}
Summing \eqref{1a} and \eqref{1b} we get the final result
\begin{eqnarray}\label{c26}
H_4| Z_1 {\bar Z}_1 Z\rangle=-\frac{1}{16}g^4| Z {\bar Z}_2 Z_2\rangle,\,\,\,c_7=-\frac{1}{16}g^4.
\end{eqnarray}

\begin{figure}[t]
  \centering
  \includegraphics[width=0.6\textwidth]{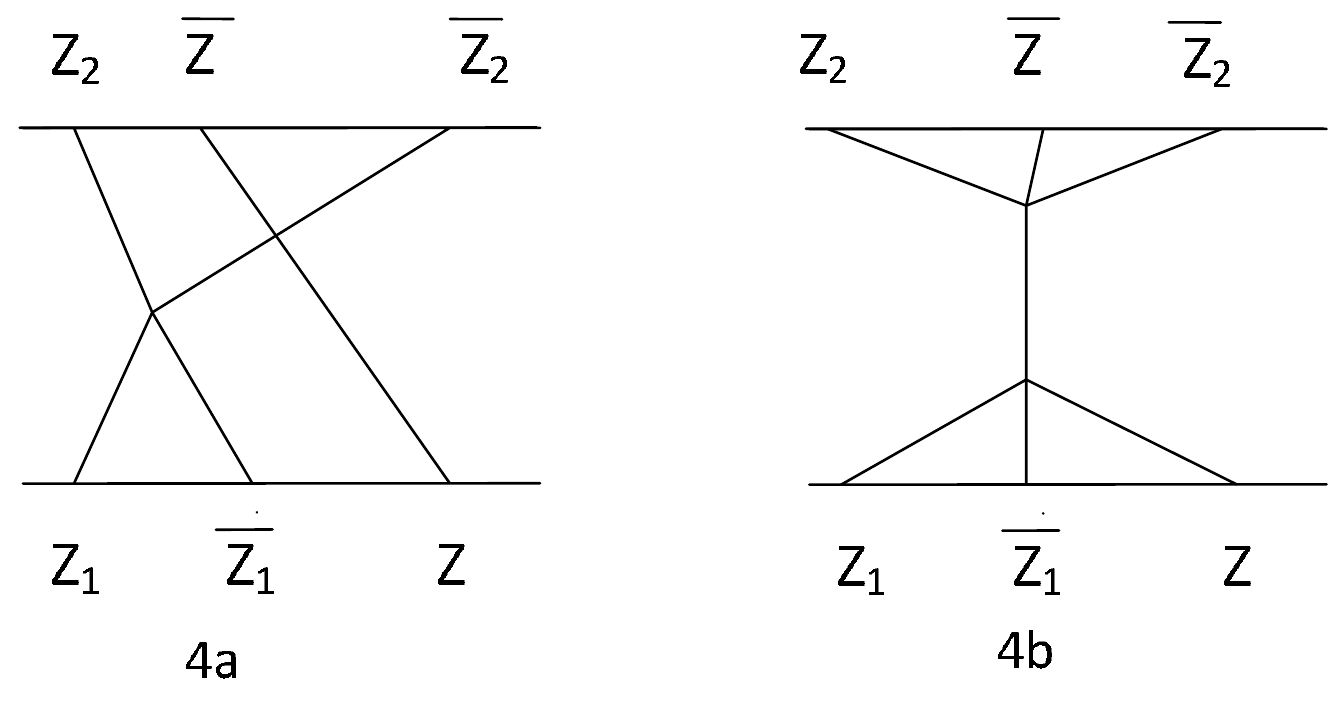}
  \caption{Diagrams contributing to the coefficient $c_8$. The bottom vertex of 4a is $-\frac{1}{2}$ times the bottom vertex of 3c
  while the top vertex of 4a is equal to the top vertex of 3c. The value of the diagram 4b can be obtained from the result of Appendix A. }
  \label{fig:current1}
\end{figure}

Next we evaluate the coefficient $c_8$. To this end we draw the relevant diagrams in Figure 4.
Again by looking at \eqref{4vertex} one can verify that the diagram 4a  is $-\frac{1}{2}$
the diagram 3c. The contribution of the 'chicken' diagram 4b can be, as before, obtained from
Appendix A.
Thus we can write
\begin{eqnarray}\label{2a}
H_4^{(4a)}| Z_1 {\bar Z}_1 Z\rangle=-\frac{1}{2}\,\,\, \frac{-1}{2}g^4|  {\bar Z}_2 Z Z_2\rangle
=\frac{1}{4}g^4|{\bar Z}_2 Z Z_2 \rangle.
\end{eqnarray}
The contribution of diagram 4b to $Z^A_{\, \, \,B}$ reads
\begin{eqnarray}\label{1bZ}
\cZ^{A(4b)}_{\, \, \,B}=-\frac{(g_{YM}^2 \mu^{2 \epsilon} N)^2}{16 (8 \pi^2)^2} \frac{1}{2 \epsilon},
\end{eqnarray}
from which one can deduce the contribution to the corresponding  anomalous dimension matrix element to be
\begin{eqnarray}\label{2b}
H_4^{(4b)}| Z_1 {\bar Z}_1 Z\rangle=-\frac{1}{8}g^4| {\bar Z}_2 Z Z_2\rangle,
\end{eqnarray}
Summing \eqref{2a} and \eqref{2b} we get the final result
\begin{eqnarray}\label{c25}
H_4| Z_1 {\bar Z}_1 Z\rangle=\frac{1}{8}g^4|{\bar Z}_2 Z Z_2 \rangle,\,\,\,c_8=\frac{1}{8}g^4.
\end{eqnarray}
\begin{figure}[t]
  \centering
  \includegraphics[width=0.25\textwidth]{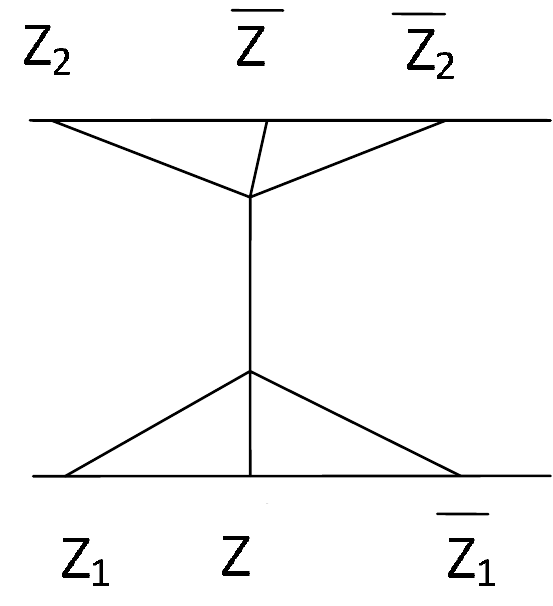}
  \caption{Diagram contributing to the coefficient $c_6$. }
  \label{fig:current1}
\end{figure}

The next coefficient to be evaluated is $c_6$.
This can be achieved by focusing on the following matrix element
\begin{eqnarray}\label{Z1ZZ1b}
H_4| Z_1 Z {\bar Z}_1 \rangle=c_6 |{\bar Z}_2 Z Z_2 \rangle+... ,
\end{eqnarray}
where the dots in the right hand side of \eqref{Z1ZZ1b} denote terms that dot not
involve the state $|{\bar Z}_2 Z Z_2 \rangle$.
The contribution to this coefficient comes from a single diagram depicted in Figure 5.
\begin{eqnarray}\label{3}
\cZ^{A(5)}_{\, \, \,B}=\frac{(g_{YM}^2 \mu^{2 \epsilon} N)^2}{8 (8 \pi^2)^2} \frac{1}{2 \epsilon},
\end{eqnarray}
from which one can deduce the contribution to the corresponding  anomalous dimension matrix element to be
\begin{eqnarray}\label{f25}
H_4| Z_1 Z {\bar Z}_1 \rangle=\frac{1}{4}g^4|{\bar Z}_2 Z Z_2 \rangle,\,\,\,c_6=\frac{1}{4}g^4.
\end{eqnarray}

We close this Section by finding the value of the last undetermined coefficient $c_5$.
Unfortunately in this case the number of diagrams proliferates and the direct diagrammatic
method used in the above is not as easy as before. Fortunately however, one can
determine $c_5$ by demanding that the operator
${\cal O}_3=\tr{\Phi_i \Phi_i Z}$ has the known
2-loop anomalous dimension $-6 g^4$ as was determined in the work of \cite{Beisert:2003tq}
by computing the anomalous dimension of a descendant in the $SU(2)$ sector.
At this point we should mention that the operator ${\cal O}_3$
does not mix with operators including fermions and covariant derivatives \cite{Georgiou:2009tp}
and as such it is ideal for using it to determine the last unknown coefficient.
To this end we evaluate the action of $H_4$ on ${\cal O}_3$.
One gets
\begin{eqnarray}\label{O3fin}
H_4 {\cal O}_3=8 (2 c_5+c_6+2(c_7+2c_8))g^4 {\cal O}_3.
\end{eqnarray}
From \eqref{O3fin} it is possible to find the last unknown coefficient $c_5$ by using
the fact that the 2-loop anomalous dimension of ${\cal O}_3$ is equal to $-6 g^4$.
Plugging in \eqref{O3fin} the values for $c_6=\frac{1}{4}g^4$, $c_7=-\frac{1}{16}g^4$ and
 $c_8=\frac{1}{8}g^4$ we get
$c_5=-\frac{11}{16}g^4$.

We are now in position to write down the 2-loop planar Hamiltonian in the $SO(6)$ sector.
It is given by
\begin{align}\label{Hamfinal}
H_{4}&= g^4 \, \Bigl (
-2\, \spinchain{abc}{abc}
+ \frac{3}{2}\, \Bigl [ \spinchain{bac}{abc} +  \spinchain{acb}{abc}\Bigr ]
-\frac{1}{2} \, \Bigl [ \spinchain{bca}{abc} +  \spinchain{cab}{abc}\Bigr ] \nn\\
& \qquad -\frac{11}{16} \Bigl [ \spinchain{bbc}{aac} +  \spinchain{cbb}{caa}\Bigr ]
+ \frac{1}{4}\, \spinchain{bcb}{aca}
-\frac{1}{16}\, \Bigl [ \spinchain{bbc}{caa} +  \spinchain{cbb}{aac}\Bigr ]\nn\\
& \qquad +\frac{1}{8}\, \Bigl [ \spinchain{bbc}{aca} +  \spinchain{cbb}{aca}
+\spinchain{bcb}{aac} +  \spinchain{bcb}{caa} \Bigr ]\, .
\end{align}

\section{Resolving the mixing among primary operators}
In this Section, we discuss the resolution of the mixing among primary operators up to order $g^2$
using the expression for the scalar piece of the 2-loop dilatation operator of $\mathcal{N}=4$ SYM \eqref{Hamfinal}.
The knowledge of the exact form of the eigenstates up to order $g^2$
is crucial for the 1-loop computation of three-point correlators involving primary operators,
which we will report upon in a
companion paper \cite{us2}.

The starting point is a set of primary operators ${\cal O}_{p_i},\, i=1,...,n$ which have the same quantum numbers and
naive dimension and diagonalize the 1-loop Hamiltonian $H_2$ of \cite{Minahan:2002ve}.
Here we will make the simplifying assumption that
the one loop eigenvalues are non-degenerate.
It is known that these 1-loop eigenstates mix with other operators
having fermions and covariant derivatives.
The resolution of the mixing with fermions can be achieved using the method of \cite{Georgiou:2008vk,Georgiou:2009tp}
or equivalently by diagonalizing $H_2+H_3$ up to order $g^3$ \cite{Xiao:2009pv}.
Let us denote this eigenstate of $H_2+H_3$ by ${\cal O}_{p_i}+ {\cal O}_{\psi_i}$.
What we are after is the next $g^2$ correction to the form of the primary operator.
This correction will include an operator whose letters are scalar fields and can be written as  a linear combination of the
1-loop eigenstates ${\cal O}_{p_i},\, i=1,...,n$. One way to determine it is the following.
Firstly, we seek the operator which diagonalizes $H_2+H_3$ not up to order $g^3$ but up to order $g^4$.
In other words we look for a solution to the eigenvalue problem
\begin{eqnarray}\label{exacteigen}
(H_2+H_3)(|{\cal O}_{p_i}\rangle+ |{\cal O}_{\psi_i}\rangle+|{\cal O}_{e_i}\rangle)=(E_{2i}+{\tilde E}_{4i})(|{\cal O}_{p_i}\rangle+ |{\cal
O}_{\psi_i}\rangle+|{\cal O}_{e_i}\rangle)
\end{eqnarray}
up to order $g^4$, where ${\cal O}_{e_i}$ is a scalar operator of order $g^2$. Similarly ${\tilde E}_{4i}$, $E_{2i}$ and ${\cal O}_{\psi_i}$ are of order $g^4$, $g^2$
and $g$ respectively.
As commented above \eqref{exacteigen} is satisfied up to order $g^3$ by definition.
At order $g^4$ \eqref{exacteigen} gives
\begin{eqnarray}\label{g^4}
H_2|{\cal O}_{e_i}\rangle+H_3 |{\cal
O}_{\psi_i}\rangle=E_{2i} |{\cal O}_{e_i}\rangle+{\tilde E}_{4i}|{\cal O}_{p_i}\rangle.
\end{eqnarray}
On general grounds, the scalar operator ${\cal O}_{e_i}$ can be written as a linear combination
of the 1-loop eigenstates. Namely,
\begin{eqnarray}\label{Oei}
{\cal O}_{e_i}=\sum_{j\neq i} c_j^{i}\,\, {\cal O}_{p_j}, \,\,\,c_j^{i}\sim O(g^2).
\end{eqnarray}
By taking the product of \eqref{g^4}  with the ket $\langle {\cal O}_{p_m}|,\,\,\,m\neq i$ and using
the orthonormality of the operators $ {\cal O}_{p_i}$
we obtain the values for the coefficients $c_m^{i}$. These are given by
\begin{eqnarray}\label{cmi}
 c_m^{i}=\frac{\langle{\cal O}_{p_m}|H_3|{\cal O}_{\psi_i}\rangle}{E_{2i}-E_{2m}} ,\,\,\,m\neq i.
\end{eqnarray}
The projection of  \eqref{g^4} along $\langle {\cal O}_{p_i}|$
will give the value of ${\tilde E}_{4i}$ .
\begin{eqnarray}\label{E4i}
{\tilde E}_{4i}= \langle{\cal O}_{p_i}|H_3|{\cal O}_{\psi_i}\rangle.
\end{eqnarray}
The complete resolution of the mixing up to order $g^2$ can be achieved by
considering the 2-loop Hamiltonian obtained in the previous Section as a perturbation to
the Hamiltonian of \eqref{exacteigen} and making use of time independent non-degenerated
perturbation theory. The correction to the eigenstate and eigenvalue coming from $H_4$
should then be added to the eigenstate and eigenvalue of \eqref{exacteigen} to obtain
the final 2-loop energy eigenvalue
\begin{eqnarray}\label{E4ifinal}
 E_{4i}= \langle{\cal O}_{p_i}|H_3|{\cal O}_{\psi_i}\rangle+\langle{\cal O}_{p_i}|H_4|{\cal O}_{p_i}\rangle.
\end{eqnarray}
 and its corresponding eigenstate
 \begin{eqnarray}\label{eigenstatefinal}
{\tilde{\cal O}}_{p_i}={\cal O}_{p_i}+ {\cal O}_{\psi_i}+ \sum_{m\neq i}\frac{\langle{\cal O}_{p_m}|H_3|{\cal O}_{\psi_i}\rangle+
\langle{\cal O}_{p_m}|H_4|{\cal O}_{p_i}\rangle}{E_{2i}-E_{2m}}\,{\cal O}_{p_m}.
\end{eqnarray}
Notice that in \eqref{eigenstatefinal} we have not written the
$O(g^2)$ single trace operators having covariant derivatives. These terms originate from the
non-diagonal elements of $H_4$ between scalar operators and operators having derivatives
and do not alter the 2-loop eigenvalue of the energy \eqref{E4ifinal}.
They can be evaluated as in \cite{Georgiou:2009tp}.

Finally, let us mention that we have performed an independent check of
\eqref{Hamfinal} by evaluating through \eqref{E4ifinal} the 2-loop anomalous dimension
of the primary operator whose leading term is $\sum_{p=0}^2\cos{\frac{\pi  (2p+3)}{2+3}}
\tr{\Phi_{AB} Z^p \Phi^{AB} Z^{2-p}}$ to find perfect agreement with the 2-loop anomalous dimension
of a level four descendant of this primary that belongs in an $SU(2)$ subsector \cite{Beisert:2003tq}.

\vspace{1cm}

\noindent {\large {\bf Acknowledgments}}

\vspace{3mm}

\noindent
We wish to thank Johannes Henn, Rodolfo Russo and Christoph Sieg
for useful discussions and comments. This research was supported in part by the Volkswagen-Foundation
and by the United States National Science Foundation under Grant No.~NSF PHY05-51164. J.P.~thanks
the KITP, Santa Barbara, for hospitality where this work was completed.

\appendix

\section{Appendix A}
In this Appendix we evaluate the "chicken" diagram of Figure 3b in dimensional regularization.
Here we chose to work with Euclidean $\mathcal{N}=4$ SYM.
\begin{eqnarray}\label{H}
H=\frac{(2 g_{YM}^2  \mu^{2 \epsilon})^2 N^4}{2^7}\int d^{2 \omega}y_2 d^{2 \omega}y_1 \Delta^3(y_2-x_2)\Delta(y_1-y_2)\Delta^3(y_1-x_1)
\end{eqnarray}
where the scalar propagator is
\begin{eqnarray}\label{propagator}
\Delta(x)=\frac{\Gamma(\omega-1)}{4 \pi^{\omega}}\frac{1}{x^{2(\omega-1)}}
\end{eqnarray}
For the sake of simplicity we have not written explicitly the free $SU(N)$ indices
of the operators appearing in the diagram 3b.

First we perform the $y_1$ integration which gives a finite result
\begin{eqnarray}\label{I1}
I_1=\int d^{2 \omega}y_1 \Delta(y_1-y_2)\Delta^3(y_1-x_1)= \nonumber \\
\Big(\frac{\Gamma(\omega-1)}{4 \pi^{\omega}}\Big)^4
\frac{\pi^{\omega}\Gamma(4(\omega-1)-\omega)}{\Gamma(\omega-1)\Gamma(3(\omega-1))}\frac{1}{-1+2 \epsilon}
(y_2-x_1)^{2(-3 \omega+4)},\,\,\,\omega=2-\epsilon.
\end{eqnarray}
Subsequently we perform the second integration
\begin{eqnarray}\label{I2}
I_2=\int d^{2 \omega}y_2 \frac{1}{(y_2-x_2)^{2(3(\omega-1))}} \frac{1}{(y_2-x_1)^{2(3\omega-4)}}= \nonumber \\
\frac{\pi^{\omega}\Gamma(7(\omega-1)-2 \omega)}{\Gamma(4(\omega-1)-\omega)\Gamma(3(\omega-1))}
\frac{\Gamma(4-2 \omega)\Gamma(3-2 \omega)}{\Gamma(7-4 \omega)}
(x_1-x_2)^{2(2 \omega-7(\omega-1))}.
\end{eqnarray}
Putting everything together and keeping in mind that $\omega=2-\epsilon$ it is
straightforward to verify that the diagram of Figure 1b has a simple pole in $\epsilon$.
\begin{eqnarray}\label{Hfin}
H=\frac{N^2}{2^3}\frac{1}{(4 \pi^2)^3}\frac{1}{(x_{12}^2)^3}\big(- \frac{(g_{YM}^2 \mu^{2 \epsilon}N)^2}{32(8 \pi^2)^2} \frac{1}{ \epsilon}    \big)
+O(\epsilon^0).
\end{eqnarray}
We should notice the we have factored out three propagators (the three first fractions in \eqref{Hfin}).
Consequently, one needs to include an appropriate term in the expression for $Z$.
This terms should be $- \frac{1}{2}$ the value of the parenthesis of \eqref{Hfin} and  is
\begin{eqnarray}\label{Zcomp}
\cZ^{A(1b)}_{\, \, \,B}=\frac{(g_{YM}^2 \mu^{2 \epsilon} N)^2}{32 (8 \pi^2)^2} \frac{1}{2 \epsilon}.
\end{eqnarray}
This factor of $- \frac{1}{2}$ is due to the fact that in \eqref{finitecorr.2} the correlator
$\langle \tilde {\cal O}^{\dagger}_{B}(x_2) \, \, \, \tilde {\cal O}^A(x_1)\rangle$ appears
just once while ${\tilde \cZ}^A_{\, \, \,B}$ appears twice, once in the second and once in the third
line of \eqref{finitecorr.2}. The same $- \frac{1}{2}$ factor is necessary in order to reproduce
the correct 1-loop anomalous dimension of \cite{Minahan:2002ve}.

\bibliographystyle{nb}
\bibliography{botany}

\end{document}